\documentclass{article}
\usepackage{spconf,amsmath,graphicx}
\usepackage{booktabs}
\usepackage{caption}

\title{The NPU-ASLP System for Audio-Visual Speech Recognition \\in MISP 2022 Challenge}
\name{
\begin{tabular}{c}
  \it Pengcheng Guo$^{\dag}$, He Wang$^{\dag}$, Bingshen Mu, Ao Zhang, Peikun Chen
\end{tabular} \thanks{\dag Equal contribution.}
}
\address{
  Audio, Speech and Language Processing Group (ASLP@NPU), School of Computer Science, \\Northwestern Polytechnical University, Xian, China\\
}
\begin{document}
\ninept
\maketitle
\begin{abstract}
\vspace{-0.1cm}
This paper describes our NPU-ASLP system for the Audio-Visual Diarization and Recognition (AVDR) task in the \textbf{M}ulti-modal \textbf{I}nformation based \textbf{S}peech \textbf{P}rocessing (MISP) 2022 Challenge.
Specifically, the weighted prediction error (WPE) and guided source separation (GSS) techniques are used to reduce reverberation and generate clean signals for each single speaker first.
Then, we explore the effectiveness of Branchformer and E-Branchformer based ASR systems.
To better make use of the visual modality, a cross-attention based multi-modal fusion module is proposed, which explicitly learns the contextual relationship between different modalities.
Experiments show that our system achieves a concatenated minimum-permutation character error rate (cpCER) of 28.13\% and 31.21\% on the Dev and Eval set, and obtains a second place in the challenge.
\end{abstract}
\begin{keywords}
Multimodal, Audio-Visual Speech Recognition
\end{keywords}

\vspace{-0.2cm}
\section{Introduction} \label{sec:intro}
\vspace{-0.2cm}
With the advances of deep learning, lots of progress has been achieved for automatic speech recognition (ASR) and its performance has been improved significantly.
However, ASR systems are still susceptible to performance degradation in real-world far-filed scenarios like meetings or home parties, due to the background noise, inevitable reverberation, and multiple speakers overlapping.
To achieve a robust ASR system in such challenging acoustic environments, plenty of studies focus on combining a separate speech enhancement module with the ASR model or an end-to-end optimization of all components.
In addition, audio-visual based ASR (AV-ASR) has also drawn immense interest when both auditory and visual data are available, since the additional visual cues, such as facial/lip movements, could provide complementary information and increase the model's robustness, especially in noisy conditions.

Inspired by this, the first Multi-modal Information based Speech Processing (MISP) Challenge~\cite{chen2022misp, 2022misptask2} was launched, which targeted exploring the usage of both audio and video data in distant multi-microphone conversational wakeup and recognition tasks.
Different from the first MISP Challenge that provides oracle speaker diarization results, this year, the MISP 2022 Challenge removes such prior knowledge and extends previous tasks to more generic scenarios, which are audio-visual speaker diarization (AVSD), and audio-visual diarization and recognition (AVDR).

This study describes our system for the AVDR task (Task2) of the MISP 2022 Challenge.
To develop a robust AV-ASR system, we first explore several commonly used data processing techniques, including the weighted prediction error (WPE)~\cite{yoshioka2012generalization} based dereverberation, guided source separation (GSS)~\cite{boeddeker2018front} and data simulation.
Then, advanced end-to-end architectures like Branchformer~\cite{peng2022branchformer} and E-Branchformer~\cite{kim2023branchformer} are used to build the basic ASR systems with the joint connectionist temporal classification (CTC)/attention training.
To better make use of the visual modality, we propose a cross-attention based multi-modal fusion module, which explicitly learns the contextual relationship between different modalities.
After combining the results from various systems by the Recognizer Output Voting Error Reduction (ROVER) technique, we achieve a final concatenated minimum permutation character error rate (cpCER) of 28.13\% and 31.21\% on the Dev and Eval set, obtaining a second place in the challenge.

\vspace{-0.2cm}
\section{Proposed System} \label{sec:method}
\vspace{-0.2cm}

\subsection{Data Processing and Simulation} \label{subsec:data}
\begin{figure}[tbp]
    \centering
    \includegraphics[width=0.8\linewidth]{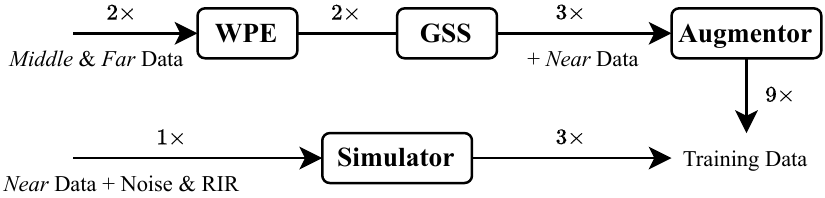}
    \vspace{-0.2cm}
    \captionsetup{font={footnotesize}}
    \caption{The flow chart of data processing and simulation. $\text{N}\times$ refers to $\text{N}$ times the original 106 hours \textit{Near} data provided by the challenge.}
    \label{fig:data_flow_chart}
    \vspace{-0.6cm}
\end{figure}

\vspace{-0.1cm}
Fig.~\ref{fig:data_flow_chart} shows our data processing and simulation progress.
Both \textit{Middle} and \textit{Far} data are first pre-processed by the WPE~\cite{yoshioka2012generalization} and GSS~\cite{boeddeker2018front} algorithms to obtain the enhanced clean signals of each speaker.
Then, an Augmentor module conducts speech perturbation on the combination of enhanced data and original \textit{Near} data, resulting in about 9-fold training data\footnote{N-fold data refer to N times the original 106 hours \textit{Near} data.}.
For the simulation part, the MUSAN corpus\footnote{https://www.openslr.org/17/} and the open-source pyroomacoustics toolkit\footnote{https://github.com/LCAV/pyroomacoustics} are applied to generate background noises and room impulse responses (RIRs). 
The total training data is about 1300 hours.

\vspace{-0.4cm}
\subsection{Audio based Speech Recognition} \label{subsec:asr}
\vspace{-0.1cm}
For the audio based ASR systems, we investigate the effectiveness of the recently proposed Branchformer~\cite{peng2022branchformer} and E-Branchformer~\cite{kim2023branchformer} architectures.
The Branchformer encoder adopts two parallel branches to capture various ranged contexts.
While one branch employs self-attention to learn long-range dependencies, the other branch utilizes a multi-layer perceptron module with convolutional gating (cgMLP) to extract fine-grained local correlations synchronously.
In~\cite{kim2023branchformer}, Kim \textit{et al}. enhanced Branchformer by applying a depth-wise convolution based merging module and stacking an additional pointwise feed-forward module, named E-Branchformer.

\vspace{-0.4cm}
\subsection{Audio-Visual based Speech Recognition} \label{sbusec:avsr}
\begin{figure}[tbp]
    \centering
    \includegraphics[width=0.6\linewidth]{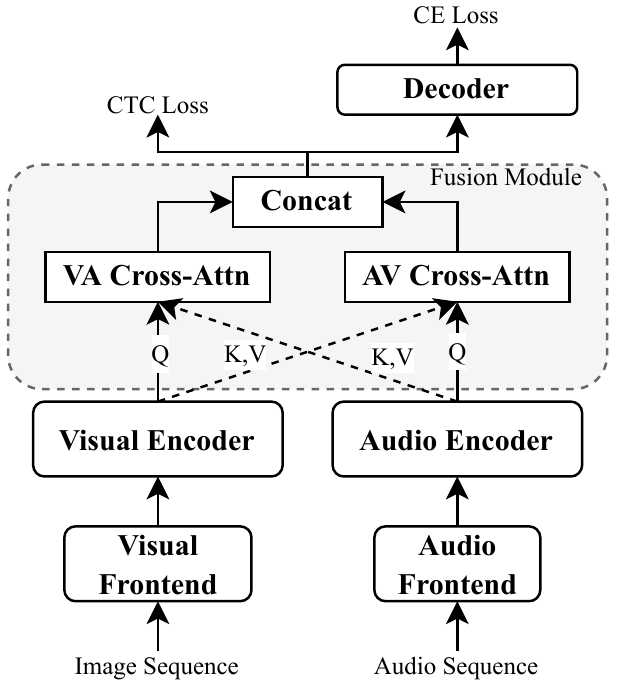}
    \vspace{-0.2cm}
    \captionsetup{font={footnotesize}}
    \caption{An overview of the proposed AV-ASR model.}
    \vspace{-0.1cm}
    \label{fig:avsr}
    \vspace{-0.2cm}
\end{figure}

\vspace{-0.1cm}
Fig.~\ref{fig:avsr} shows the overall framework of our AV-ASR model.
In detail, each modality is first processed by a frontend module to extract features. The visual frontend is a 5-layer ResNet3D module, while the audio frontend is a 2-layer convolutional subsampling module.
Following the frontends, two modal-dependent Branchformer encoders are used to encode input features as latent representations.
The proposed fusion module consists of 2 cross-attention layers, each of which takes one modality as the Query vector and the other modality as Key/Value vectors. With the help of cross-attention layers, each modality could learn the related and complementary context from the other modality.
Finally, representations from different modalities are concatenated together to compute CTC loss and fed into the Transformer decoder to compute cross-entropy (CE) loss.

\vspace{-0.4cm}
\subsection{Inference Procedure} \label{subsec:overall_framework}
\vspace{-0.1cm}
During the inference, the Eval set is first segmented by a speaker diarization (SD) model, enhanced by WPE and GSS, transcribed by our ASR or AV-ASR models, and rescored by a Transformer based language model (LM).
Our SD model is implemented based on the released baseline system\footnote{https://github.com/mispchallenge/misp2022\_baseline} by replacing the long short-term memory (LSTM) based diarization module with Transformer and the speaker embedding module with an ECAPA model pre-trained on all available speaker recognition and verification data in the challenge homepage\footnote{https://mispchallenge.github.io/mispchallenge2022/extral\_data.html}.
The diarization error rates (DERs) of baseline SD and our SD are 13.09\% and 9.43\% on the Dev set, respectively.
Finally, results from different systems are fused by the ROVER technique.


\vspace{-0.2cm}
\section{Experiments} \label{sec:experiments}
\vspace{-0.2cm}
\subsection{Setup} \label{subsec:setup}
\vspace{-0.1cm}
All of the models are implemented with ESPnet~\cite{watanabe2018espnet}.
For the audio based ASR systems, we follow the ESPnet recipe to set the framework: Enc = 24, Dec = 6.
Since the additional modules increase the parameters of E-Branchformer, we also train an E-Branchformer Small (Enc = 16) for a fair comparison.
For AV-ASR systems, the visual front is a 5-layer ResNet3D module, whose channels are 32, 64, 64, 128, 256 and kernel size is 3, the visual encoder is a 12-layer Branchformer, and others are the same as ASR systems. During the training, the audio branch is initialized by well-trained ASR models.

\vspace{-0.4cm}
\subsection{Results} \label{subsec:results}
\vspace{-0.1cm}
Table~\ref{tab:asr_results} presents the cpCER results of various ASR systems. It can be seen that all of our models gives better results over the official baseline and achieve up to 40\% absolute cpCER improvement due to the enhanced dataset and advanced model architectures. Comparing M1 and M2, the data simulation gives a noticeable gain. Besides, E-Branchfomer Small (M3) obtains similar results with Branchformer (M1) when model sizes are the same, and increasing the encoder block of E-Branchformer (M4) gives the best results. When comparing the results in each row, we find that a better SD model could bring consistent performance improvement.
Table~\ref{fig:avsr} shows the cpCER results of our AV-ASR systems.
For AV-ASR, a good initialization of the ASR branch gives better performance (M5 vs. M6). Comparing M6 and M1, the incorporation of visual modality gives about 0.9\% cpCER improvement.
After fusing systems of M1, M3, M4, M5 and M6, we obtain a cpCER of 28.13\% and 31.21\% on the Dev and Eval set (w/ our SD), achieving second place in the challenge.

\begin{table}[tbp]
\centering
\captionsetup{font={footnotesize}}
\caption{The cpCER (\%) results of ASR systems on the Dev set. The Dev set is segmented by oracle timestamps, baseline SD model, and our SD model.}
\vspace{-0.3cm}
\label{tab:asr_results}
\resizebox{1.0\linewidth}{!}{
    \begin{tabular}{c|l|c|c|c} 
    \toprule[1.5pt]
    \textbf{Sys.} & \textbf{Model} & \textbf{Oracle Timestamps} & \textbf{Base SD (DER=13.09\%)} & \textbf{Our SD (DER=9.43\%)}  \\ 
    \hline\hline
    B1 & Official Baseline & 66.07 & N/A & N/A \\
    M1 & Branchformer & 26.60 & 34.04 & 30.67 \\
    M2 & ~ ~ - remove Simu Data & 27.90 & 35.00 & 31.72 \\
    M3 & E-Branchformer Small & 26.80 & 33.83 & 30.61 \\
    M4 & E-Branchformer & \textbf{26.50} & \textbf{33.73}& \textbf{30.49} \\
    \bottomrule[1.5pt]
    \end{tabular}
}
\vspace{-0.2cm}
\end{table}

\begin{table}[tbp]
\centering
\captionsetup{font={footnotesize}}
\caption{The cpCER (\%) results of AV-ASR systems on the Dev set.}
\vspace{-0.3cm}
\label{tab:avsr_results}
\resizebox{1.0\linewidth}{!}{
    \begin{tabular}{c|l|c|c} 
    \toprule[1.5pt]
    \textbf{Sys.} & \textbf{Model} & \textbf{Oracle Timestamps} & \textbf{Our SD (DER=9.43\%)}  \\ 
    \hline\hline
    M5 & AV-Branchformer (init by M2) & 26.90 & 30.85 \\
    M6 & AV-Branchformer (init by M1) & \textbf{25.70}  & \textbf{29.73} \\
    \bottomrule[1.5pt]
    \end{tabular}
}
\vspace{-0.5cm}
\end{table}

\vspace{-0.2cm}
\section{Conclusion} \label{sec:conclusion}
\vspace{-0.2cm}
In this study, we describe our system for the Task2 of the MISP 2022 Challenge. Our efforts include data processing and simulation strategies, investigation of advanced architectures, and a novel cross-attention based multi-modal fusion model. By combining various systems, we get a cpCER of 31.21\% on the final Eval set, obtaining a second place in the challenge.
\vspace{-0.2cm}

\bibliographystyle{IEEEbib}
\bibliography{strings,refs}

\end{document}